\documentclass[epj]{webofc}
\usepackage[varg]{txfonts}   
\wocname{EPJ Web of Conferences}
\woctitle{ICNFP 2015}
\newcommand{\pt}{p_{\mathrm{T}}}
\newcommand{\snn}{\sqrt{s_{\mathrm{NN}}}}

\begin{document}
\selectlanguage{english}
\title{Performance of the ALICE secondary vertex b-tagging algorithm}
%
%

\author{G. Eyyubova\inst{1,2}\fnsep\thanks{\email{Gyulnara.Eyyubova@fjfi.cvut.cz}} \and
        L. Kram\'arik\inst{1}\fnsep\thanks{\email{Lukas.Kramarik@cern.ch}} on behalf of the ALICE collaboration
}

\institute{FNSPE, Czech Technical University in Prague 
\and
           SINP MSU, Russia 
}

\abstract{%
The identification of jets originating from beauty quarks in heavy-ion collisions is important to study the properties of the hot and dense matter produced in such collisions. A variety of algorithms for b-jet tagging was elaborated at the LHC experiments. They rely on the properties of B hadrons, i.e. their long lifetime, large mass and large multiplicity of decay products. In this work, the b-tagging algorithm based on displaced secondary-vertex topologies is described. We present Monte Carlo based performance studies of the algorithm for charged jets reconstructed with the ALICE tracking system in p-Pb collisions at $\snn=5.02$ TeV. The tagging efficiency, rejection rate and the correction of the smearing effects of non-ideal detector response are presented.
}
\maketitle
\section{Introduction}
\label{intro}
Jet reconstruction provides access to the kinematics of partons produced  in hard scatterings in the initial stage of heavy-ion collisions and that later suffer energy loss in the medium via gluon radiation and elastic collisions. Reconstructed jets are therefore a powerful tool to study the properties of the quark-gluon plasma (QGP) created in heavy-ion collisions at high energy. The parton energy loss in the QGP depends on the colour charge and mass of the parton. Due to the larger colour charge factor for gluons, energy loss via gluon radiation should be larger for gluon jets than quark jets. For heavy quarks it is expected that radiative energy loss is suppressed due to coherence effects \cite{Dokshitzer,Armesto}. This should lead to a smaller modification of the transverse momentum spectra for particles containing heavy quarks than for inclusive charged hadrons. On the other hand, for heavy quarks in the QGP, collisional energy loss may be important at low transverse momentum $\pt$. The jet quenching phenomenon and the properties of produced matter can be further understood by measuring beauty jets at low jet transverse momentum in comparison with that of light-flavour jets \cite{vitev}.

The ALICE detector \cite{ALICE} is capable to discriminate jets originating from b quarks with different tagging algorithms. Here, we study the performance of the b-jet tagging algorithm based on displaced secondary-vertex topologies for p-Pb collisions at $\snn=5.02$ TeV. The p-Pb collisions serve as a test for cold nuclear effects which are important in measurements of jet quenching in Pb-Pb collisions. The study is based on Monte Carlo (MC) simulations. 

\section{Analysis details}
In this analysis we use MC with simulated responses of ALICE detector by GEANT3 \cite{GEANT} for p-Pb collisions at ${\snn} = 5.02 $ TeV. A MC simulation for pp collisions at $\sqrt{s} = 5.02 $ TeV with a Lorentz boost to get the same rapidity shift as in p-Pb collisions is used to study detector response for jets without underlying event. PYTHIA6, with tune Perugia 2011 \cite{PYTHIA}, is employed for MC pp simulations.  
The MC for p-Pb collisions is represented by a cocktail of event generators PYTHIA6 (Perugia 2011) + HIJING \cite{HIJING} which correspond to hard physics and p-Pb background, respectively.

Schematically, the analysis of b jets can be represented by the following analysis steps: track selection suitable for b-tagging, jet reconstruction from these tracks and underlying background subtraction under the jet area, tagging with a particular algorithm and, finally, corrections of jet spectrum. Corrections of the raw tagged spectrum $\displaystyle\frac{\mathrm{d}N^{\mathrm{tag}}}{\mathrm{d}\pt}$ have to be made for detector effects (unfolding) and for tagging efficiency and purity. 
In case of p-Pb collisions, unfolding also includes a correction for fluctuations of the underlying background under the jet area.

The efficiency of b-tagging, $\epsilon_\mathrm{b}$, and the purity, $f_\mathrm{b}$, are defined as the fraction of "true" b jets after tagging w.r.t total number of "true" b jets before tagging and w.r.t. total number of tagged jets, respectively. In order to estimate the efficiency and the purity, we use the "true" underlying flavour of a jet, which can be identified in MC by different labelling methods. 
The performance of the tagging algorithm is investigated via mistagging rate, which is the efficiency of mistakenly tagging as beauty a jet originating from a c-quark or a light-flavour parton. Mistagging rate studies are needed to extract sample with high purity of b jets. High purity allows to suppress the contamination of light-flavour and charm jets, which is important since the measured fraction of b jets to inclusive jets is at the level of 2-4\% (as measured by CMS experiment), both in pp collisions at $\sqrt{s} = 7$ TeV \cite{CMS_frac} and in p-Pb collisions at  $\snn = 5.02$ TeV \cite{CMS_frac_pPb}.
Below, we discuss the analysis steps in more details. 

\label{sec-1}
\subsection{Track selection}
In the MC simulations, charged tracks are reconstructed with simulated responses of Inner Tracking System (ITS) and Time Projection chamber (TPC) in a pseudorapidity region $|\eta|<0.9$ and transverse momentum $0.15<\pt<100$ GeV/$c$.  
The track selection is optimized for the reconstruction of beauty jets.
The main track cuts are the minimum number of points in the TPC (70 out of 159) and the maximum distance of closest approach to primary vertex (in $xy$-plane and in $z$ direction). Tracks are also required to have at least one point in the Silicon Pixel Detector (SPD). Because of inefficient regions in the SPD, the azimuthal distribution of these high-quality tracks is not completely uniform. This anisotropy is compensated by considering in addition tracks without reconstructed track points in the SPD. Track fit for these tracks is not constrained to the primary vertex. For the tracks used for secondary vertex reconstruction in jets, the point in SPD is required. 
 
\subsection{Jet reconstruction and background estimation}
\label{jets}
For the estimation of the background density and fluctuations in jet reconstruction, we use a similar approach as it was applied by ALICE for charged jet measurements in p-Pb collisions \cite{ALICE_jetpPb}.
Namely, the anti-$k_{\mathrm{T}}$ algorithm from the FastJet package \cite{FastJet} with resolution parameter of $R = 0.4$ is used to reconstruct charged jets. We use the CMS method \cite{CMS_bg}, which is suited for more sparse environment in p-Pb collisions w.r.t Pb-Pb collisions, for the calculation of the background density, $\rho$. The background fluctuations are calculated with a random cone approach as:
$$
\delta p_{\mathrm{T}}=\sum_i p_{\mathrm{T},i} - A_{cone}\cdot \rho,
$$
where the sum is over the track $p_{\mathrm{T}}$ in a cone and $A_{cone}$ is the cone area. For more details see \cite{ALICE_jetpPb}. Only jets with jet axis in the pseudorapidity interval $|\eta_{\mathrm{jet}}|<0.5$ are considered, which ensures that the jet is fully located in the TPC acceptance.
In this study, the background fluctuations are measured with HIJING p-Pb events with particle transport through the ALICE detector.
The background fluctuations under the jet area are found to be slightly smaller in the HIJING simulation than in the p-Pb data \cite{ALICE_jetpPb}.

\subsection{Correction for detector effects and background fluctuations}  
The jet momentum distribution is distorted by the detector mainly due to tracking inefficiency and momentum resolution. In order to correct for detector effects the detector simulation with GEANT is performed.
Reconstructed jets at particle level without detector effects and jets at the detector level after particle transport through the ALICE detector are matched and based on it the detector response matrix is built. From the inverse response matrix and the momentum distribution at detector level one finds the true jet momentum distribution.
The unfolding procedure has to be carrefully scrutinized since small perturbations in the measured distribution lead to large fluctuations in the solution. 
Several regularization methods exist. The Singular Value Decomposition (SVD) approach \cite{SVD} is used in this performance study. 

After subtracting the constant background in each event in order to correct the jet $\pt$, one has to keep in mind that the background is not necessary constant, but may differ for different jets in a given event.
This is corrected statistically (not event by event) via an unfolding technique with a background fluctuation matrix $f(\delta \pt)$, where the background fluctuations $\delta \pt$ are described in Sec.~\ref{jets}. 
The actual unfolding is done with a matrix which is a product of two matrices: detector response matrix and background fluctuation matrix.


\subsection{Secondary vertex tagging algorithm}
Due to the long life-time of B hadrons ($\approx$ 500 $\mu$m), in most cases their decay vertex is displaced from the primary vertex of a collision. This algorithm  reconstructs  secondary vertices (SV) in a jet and uses their properties to discriminate b jets among lighter flavour jets.
The tracks participating in SV reconstruction are tracks belonging to a jet with additional requirements: the point in the SPD (as mentioned above) and $p_\mathrm{T, track}>1$ GeV/$c$.
Only secondary vertices made of three tracks are considered. All combinations of three tracks, satisfying the requirement, are used to build secondary vertices in a jet. The quality of vertex reconstruction is characterized by the dispersion of the tracks in the vertex $\sigma_\mathrm{vtx}=\sqrt{d_1^2+d_2^2+d_3^2}$, where $d_{1,2,3}$ are the distances of the three tracks from SV. The sign of the SV flight distance is defined w.r.t. the jet direction, so that the signed length is $L=\lvert \vec{L'} \rvert \mathrm{sign}(\vec{L'}\cdot p_{jet})$, where  $\vec{L'}$ is the  SV position. The cuts on the most displaced SV found in the jet, namely SV dispersion, $\sigma_\mathrm{vtx}$, and the significance of flight distance in a transverse plane, $L_{xy}/\sigma_{L_{xy}}$, are used for b-jet tagging in this analysis.

\subsection{MC labelling}
\label{MC_label}
To assign the "true" flavour of reconstructed jets in the MC simulations, the MC event history is used. The labelling procedure is not unambiguous and is not strictly identical for different MC generators (it depends on parton shower implementation). Here, we treat a jet as coming from a beauty quark, if a B hadron was found in a cone $\Delta R<0.7$ around jet axis. If no beauty hadron is present, but charm hadron was found in a cone $\Delta R<0.7$, the jet is labelled as c-jet. All other jets are considered as light-flavour jets. 

\section{Performance of SV b-tagging algorithm}
\begin{figure}[h]
\centering
\sidecaption
\includegraphics[width=7cm,clip]{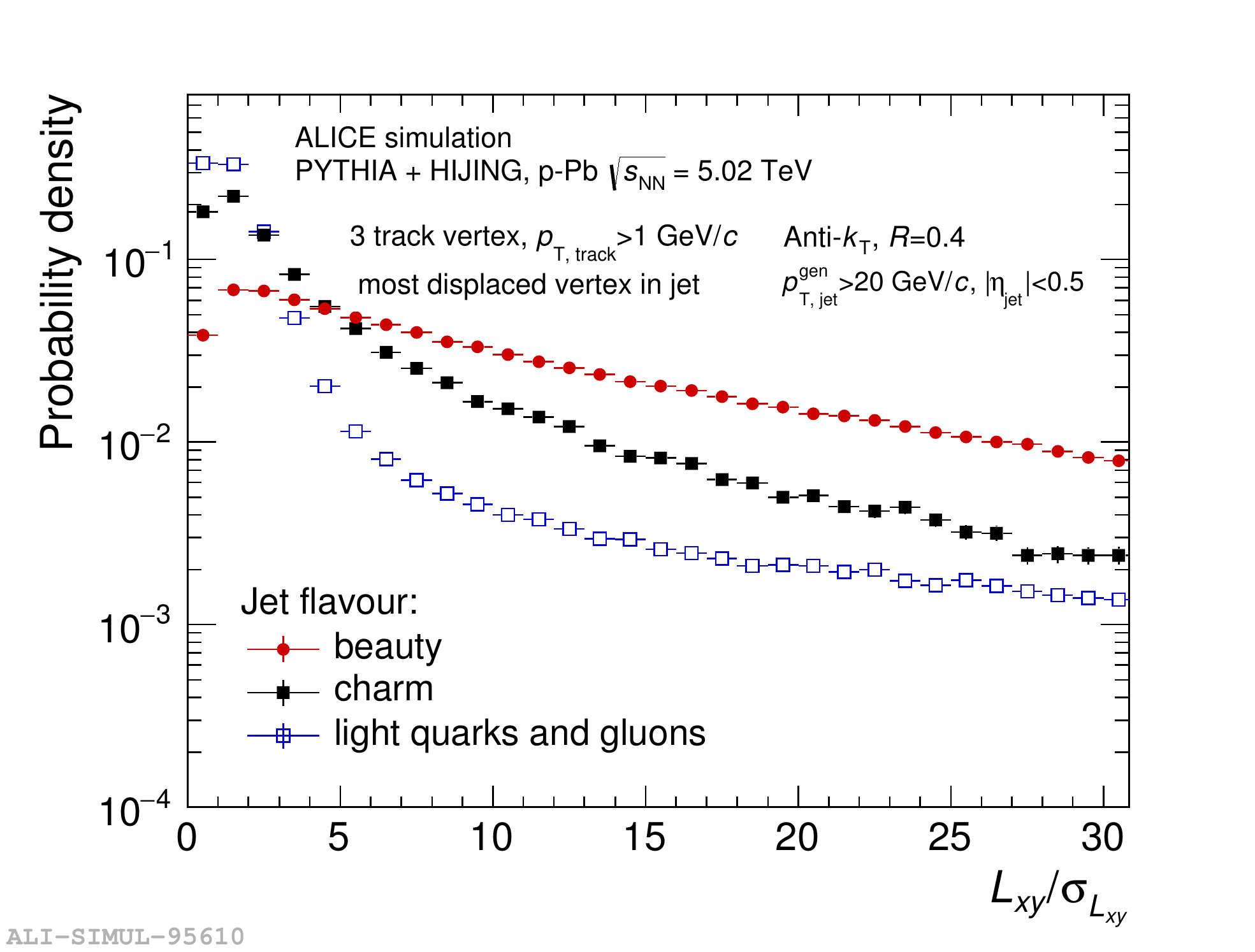}
\caption{Probability distribution of the signed flight distance significance of the most displaced secondary vertex, found in charged jets with $\pt > 20$ GeV/$c$ in p-Pb simulations at $\snn=5.02$ TeV. The different jet flavours are assigned as described in Sec.~\ref{MC_label}. }
\label{fig.SLxy}      
\end{figure}

The discrimination power of the signed flight distance significance can be judged by the distribution of this variable for jets of different flavours, which is shown in Fig.\ref{fig.SLxy}. Secondary vertices are searched in jets with $p_\mathrm{T,jet}>20$ GeV/$c$ and the most displaced vertex in a jet is considered. The larger the cut value of $L_{xy}/\sigma_{L_{xy}}$, the more light and charm jets are rejected compared to beauty jets. 
As mentioned above, the tagging procedure uses cuts on $L_{xy}/\sigma_{L_{xy}}$ and $\sigma_\mathrm{vtx}$ of SV.
Different values of cuts yield different algorithm performance. 
For a particular algorithm an operating point is defined based on inclusive b-tagging efficiency and mistagging rate.
\begin{figure}[h]
\centering
\includegraphics[width=7cm,clip]{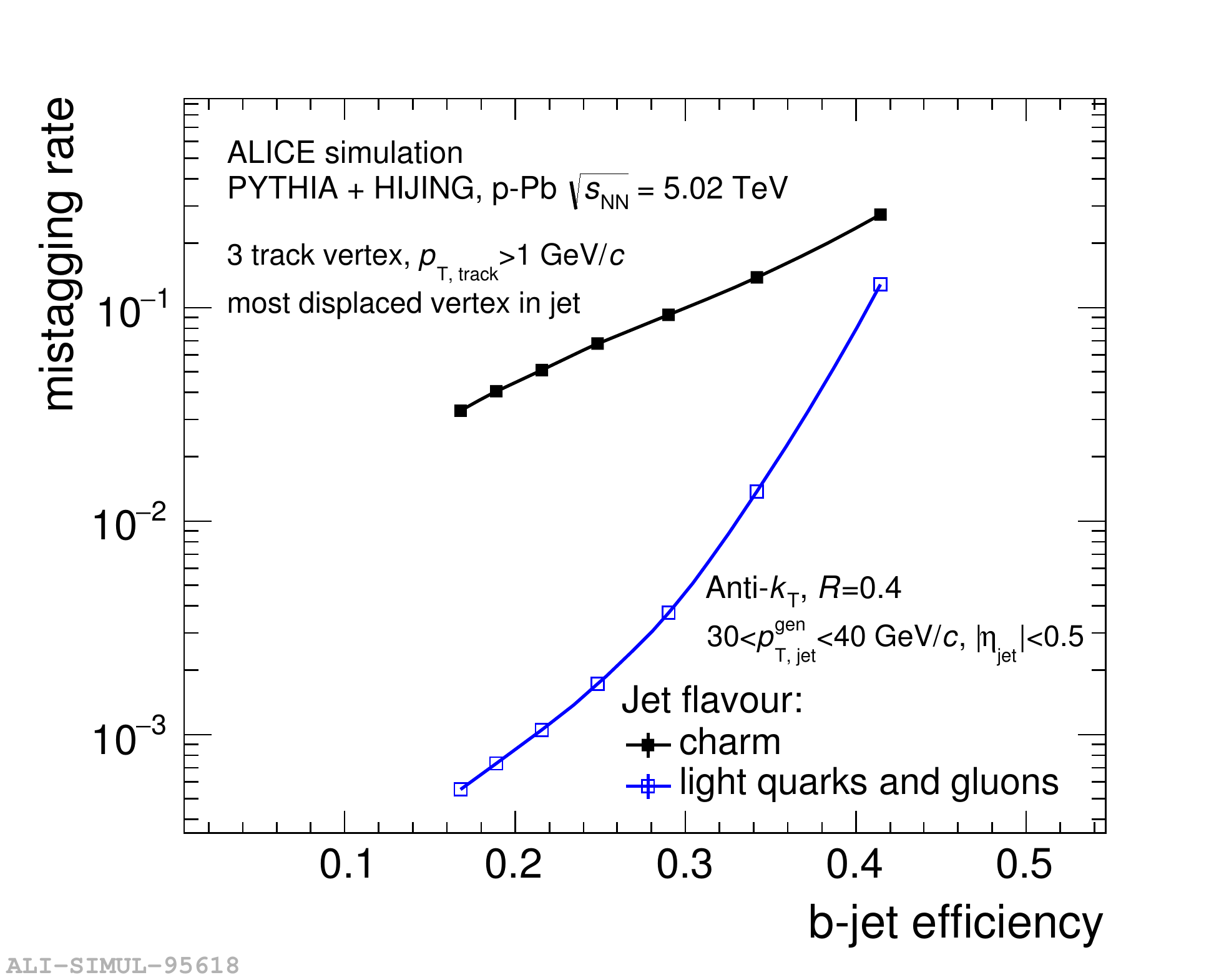}
\includegraphics[width=7cm,height=5.6cm, clip]{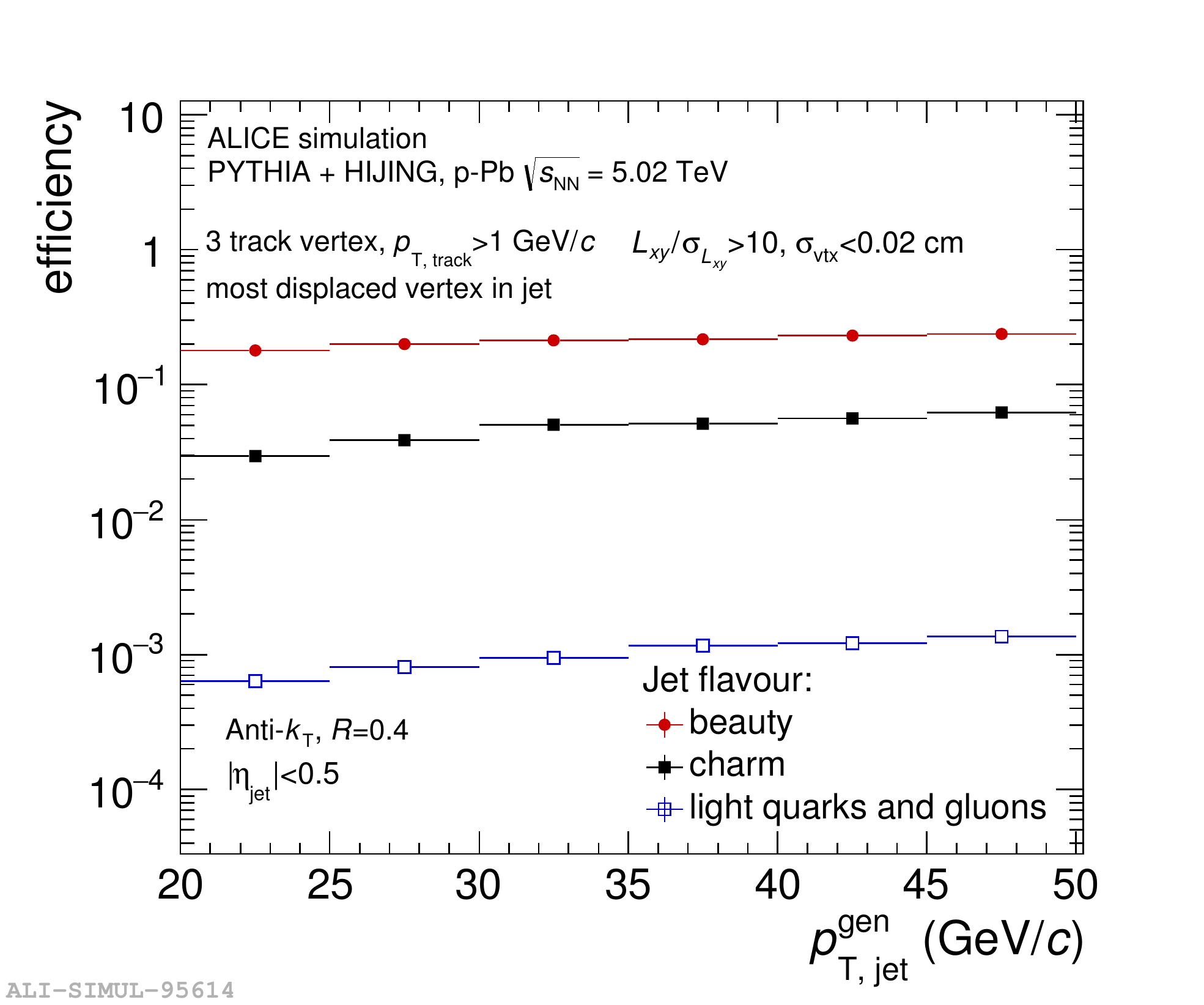}
\caption{Left: Mistagging rate vs the b-tagging efficiency of SV tagging algorithm for different operating points for $p_\mathrm{T}$ range $30<p_\mathrm{T,jet}<40$ GeV/$c$. Right: The $p^\mathrm{gen}_\mathrm{T,jet}$ dependence of the b-tagging efficiency and mistagging rate for $L_{xy}/\sigma_{L_{xy}}>10$ and $\sigma_\mathrm{vtx}<0.02$ cm.}
\label{fig.Eff}       
\end{figure}

Figure~\ref{fig.Eff}, left, shows the mistagging rate vs the b-tagging efficiency in jet $p_\mathrm{T}$ range $30<p_\mathrm{T,jet}<40$ GeV/$c$ for different operating points, which are obtained by varying the cuts on $L_{xy}/\sigma_{L_{xy}}$ (from 2 to 14), while the cut $\sigma_\mathrm{vtx}<0.02$ cm is kept fixed. The performance may depend on $p_\mathrm{T,jet}$. In the considered region, $30<p_\mathrm{T,jet}<40$ GeV/$c$, the mistagging and tagging efficiencies are almost flat. Looser cuts result in larger statistics and higher tagging efficiency, but also higher mistagging rate, and therefore reduce purity of sample. The operation point is chosen in a way that the mistagging efficiency for light-flavour jets would be about two orders of magnitude lower than the tagging efficiency. This comes from the CMS measurements on the ratio of b jets to inclusive jets, in which the b-jet fraction was found to be at the level of 2-4\% in pp collisions at $\sqrt{s} = 7$ TeV \cite{CMS_frac}.
The tagging and mistagging efficiencies $\epsilon(p^{gen}_{T,jet})$ at particle level for the chosen operating point with $L_{xy}/\sigma_{L_{xy}}>10$ and $\sigma_\mathrm{vtx}<0.02$ cm are reported in Fig.~\ref{fig.Eff}, right. The b-tagging efficiency is at the level 0.2. The efficiency to tag light-flavour jets is two orders of magnitude lower and the efficiency to tag charm jet is about 3 to 5 times lower than the b-tagging efficiency.

\begin{figure}[h]
\centering
\sidecaption
\includegraphics[width=7cm,clip]{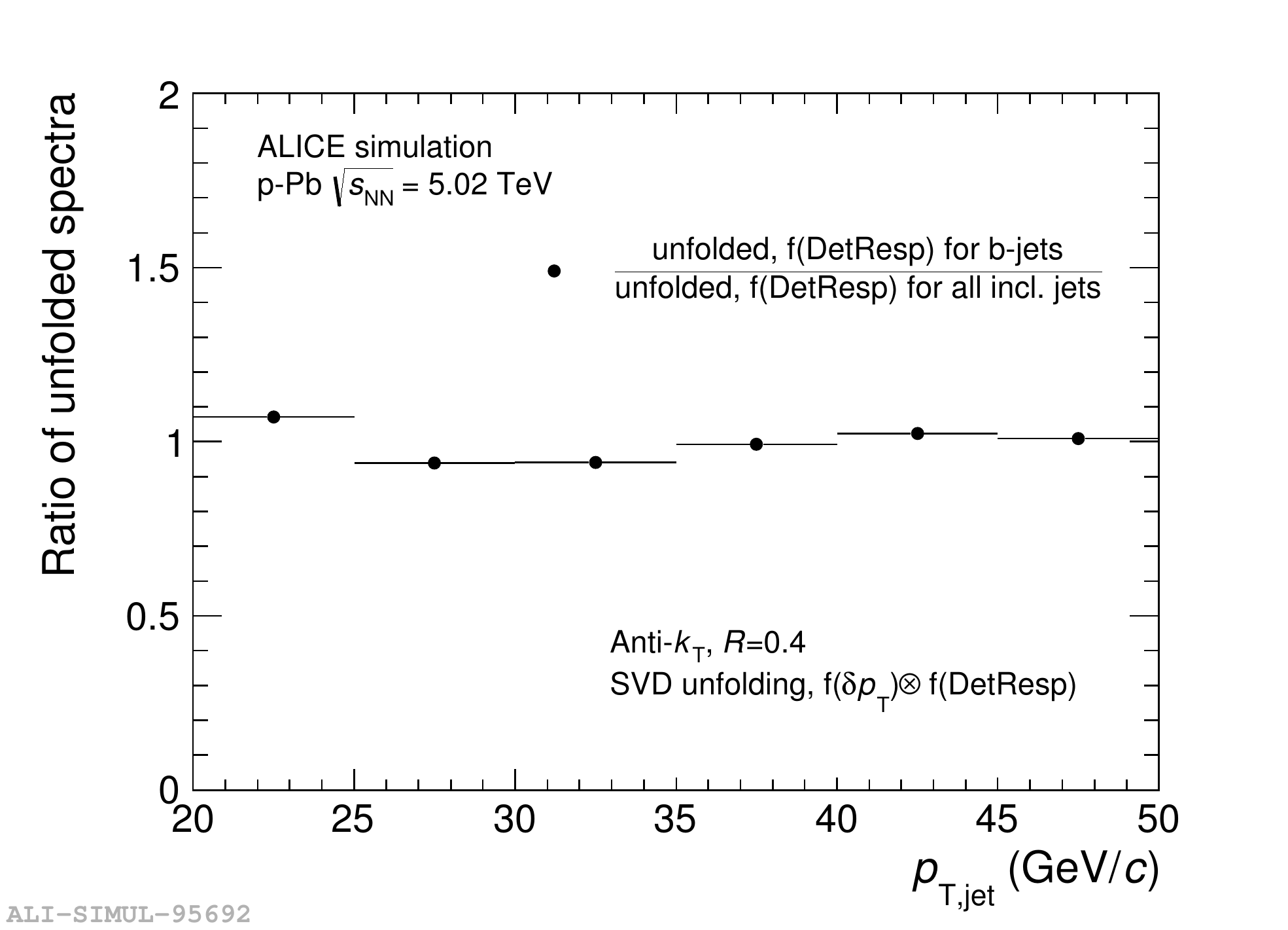}
\caption{Comparison of SVD unfolding of b-jet spectrum with two matrices: detector matrix for inclusive jets and for b jets. Both matrices are combined with background fluctuation matrix from MC. The ratio of the two unfolded results is shown.}
\label{fig.Unf1}     
\end{figure}

The correction for detector effects is tested with two detector response matrices: the detector response for b-jets and the one for inclusive jets. Figure~\ref{fig.Unf1} shows the ratio of two unfolded b-jets $\pt$-spectra. In both cases the unfolding procedure yields similar result. This ensures that beauty and inclusive jets have similar detector response matrices and therefore the detectors response for inclusive jets can be used to unfold the tagged b-jet spectrum. 
\begin{figure}[h]
\centering
\sidecaption
\includegraphics[width=7cm,clip]{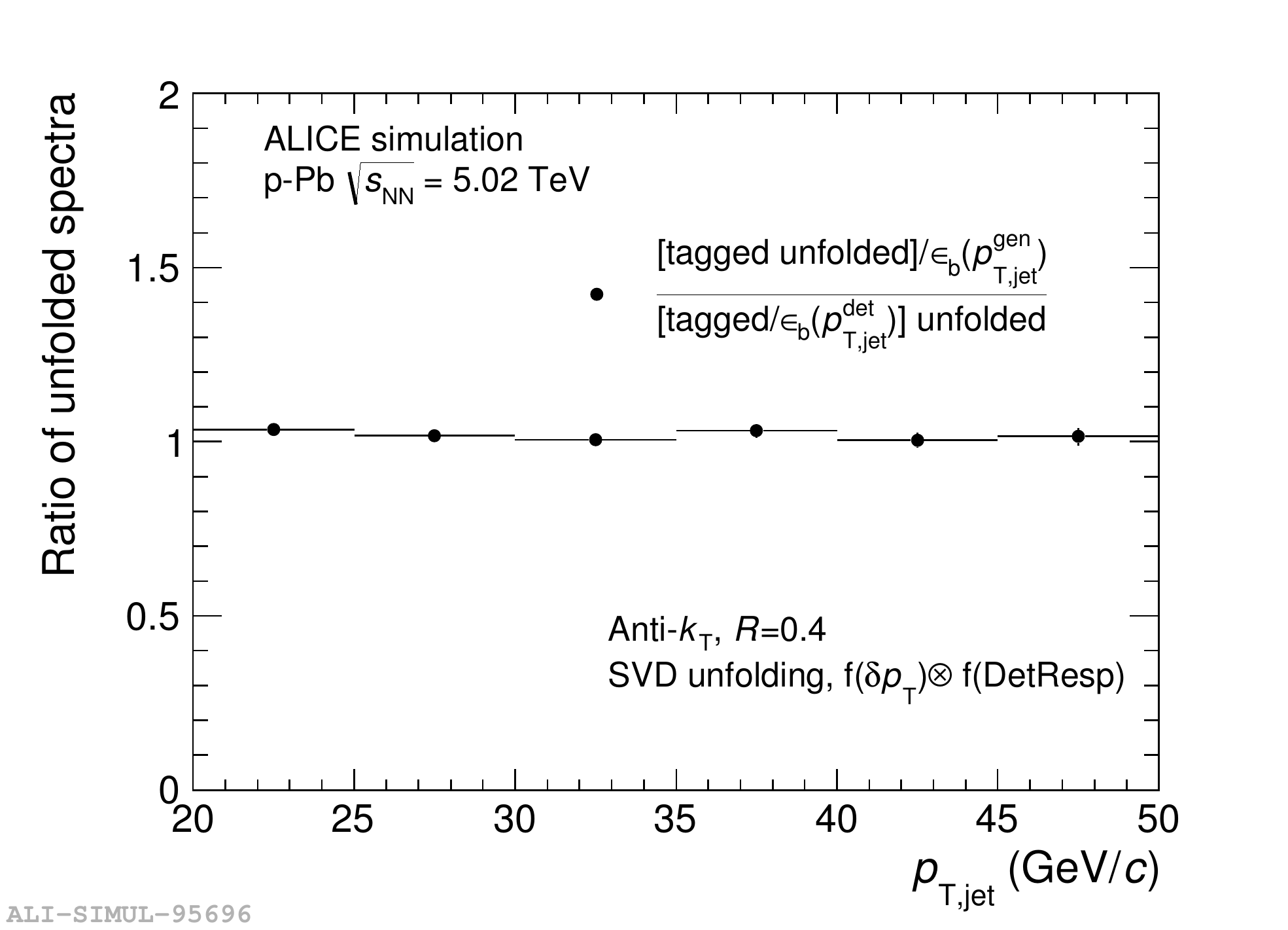}
\caption{Comparison of two sequences of corrections: SVD unfolding and correction for tagging performance and vice versa. The ratio of the two corrected spectra is shown.}
\label{fig.Unf2}       
\end{figure}

The different sequences of corrections are further studied in order to estimate the impact of background fluctuations on the stability of results. We perform the two scenarios of corrections:
\begin{enumerate}
\item The measured tagged b-jet spectrum is first unfolded for detector response and background fluctuations, then corrected for the b-tagging efficiency as a function of $p^\mathrm{gen}_\mathrm{T}$ (at particle level).
\item The measured tagged b-jet spectrum is first corrected for the b-tagging efficiency as a function of $p^\mathrm{det}_\mathrm{T}$ and then unfolded for detector response and background fluctuations.  
\end{enumerate}  

The comparison of these two scenarios is shown in Fig.~\ref{fig.Unf2} as the ratio between the two corrected spectra. The two spectra are in agreement with each other, suggesting that both scenarios can be applied in p-Pb collisions.

\section{Summary}
The performance of the b-jet tagging algorithm based on displaced secondary vertices was studied with MC simulation of p-Pb events for ALICE detector. The presented results were carried out in the jet $p_T$ range $20<p_{T,jet}<50$ GeV/$c$.

For the selected operating point the tagging efficiency is at the level 20\% while the mistagging efficiency for light-flavour jets is two orders of magnitude lower, which should result in high purity of the algorithm. The rejection of tracks with V0 topology is expected to enhance the algorithm performance by reducing light-flavour contamination and is currently under investigation.

Corrections for detector response and background fluctuations were studied.  It was found that the b-jet spectrum can be corrected with a detector response matrix for inclusive jets. The order of corrections (tagging efficiency vs unfolding) gives compatible results.

The tagging purity itself is not discussed here. Studies to estimate it via MC and data-driven approaches are ongoing.

\section{Acknowledgment}
This work was supported by the European social fund within the framework of realizing the project "Support of inter-sectoral mobility and quality enhancement of research teams at Czech Technical University in Prague", CZ.1.07/2.3.00/30.0034 and by Grant Agency of the Czech Technical University in Prague, grant No. SGS13/215/OHK4/35/14.


\end{document}